

\font\twelverm=cmr10 scaled 1200    \font\twelvei=cmmi10 scaled 1200
\font\twelvesy=cmsy10 scaled 1200   \font\twelveex=cmex10 scaled 1200
\font\twelvebf=cmbx10 scaled 1200   \font\twelvesl=cmsl10 scaled 1200
\font\twelvett=cmtt10 scaled 1200   \font\twelveit=cmti10 scaled 1200
\font\twelvesc=cmcsc10 scaled 1200
\skewchar\twelvei='177   \skewchar\twelvesy='60
\def\twelvepoint{\normalbaselineskip=12.4pt
  \abovedisplayskip 12.4pt plus 3pt minus 6pt
  \belowdisplayskip 12.4pt plus 3pt minus 6pt
  \abovedisplayshortskip 0pt plus 3pt
  \belowdisplayshortskip 7.2pt plus 3pt minus 4pt
  \smallskipamount=3.6pt plus1.2pt minus1.2pt
  \medskipamount=7.2pt plus2.4pt minus2.4pt
  \bigskipamount=14.4pt plus4.8pt minus4.8pt
  \def\rm{\fam0\twelverm}          \def\it{\fam\itfam\twelveit}%
  \def\sl{\fam\slfam\twelvesl}     \def\bf{\fam\bffam\twelvebf}%
  \def\mit{\fam 1}                 \def\cal{\fam 2}%
  \def\tt{\twelvett}
  \def\sc{\twelvesc}
  \def\nullspace{\nulldelimiterspace=0pt \mathsurround=0pt }
  \def\big##1{{\hbox{$\left##1\vbox to 10.2pt{}\right.\nullspace$}}}
  \def\Big##1{{\hbox{$\left##1\vbox to 13.8pt{}\right.\nullspace$}}}
  \def\bigg##1{{\hbox{$\left##1\vbox to 17.4pt{}\right.\nullspace$}}}
  \def\Bigg##1{{\hbox{$\left##1\vbox to 21.0pt{}\right.\nullspace$}}}
  \textfont0=\twelverm   \scriptfont0=\tenrm   \scriptscriptfont0=\sevenrm
  \textfont1=\twelvei    \scriptfont1=\teni    \scriptscriptfont1=\seveni
  \textfont2=\twelvesy   \scriptfont2=\tensy   \scriptscriptfont2=\sevensy
  \textfont3=\twelveex   \scriptfont3=\twelveex  \scriptscriptfont3=\twelveex
  \textfont\itfam=\twelveit
  \textfont\slfam=\twelvesl
  \textfont\bffam=\twelvebf \scriptfont\bffam=\tenbf
  \scriptscriptfont\bffam=\sevenbf
  \normalbaselines\rm}

\def\beginlinemode{\endmode
  \begingroup\parskip=0pt \obeylines\def\\{\par}\def\endmode{\par\endgroup}}
\def\beginparmode{\endmode
  \begingroup \def\endmode{\par\endgroup}}
\let\endmode=\par
{\obeylines\gdef\
{}}
\def\singlespace{\baselineskip=\normalbaselineskip}
\def\oneandahalfspace{\baselineskip=\normalbaselineskip
  \multiply\baselineskip by 3 \divide\baselineskip by 2}
\def\triplespace{\baselineskip=\normalbaselineskip \multiply\baselineskip by 3}
\newcount\firstpageno
\firstpageno=2
\footline={\ifnum\pageno<\firstpageno{\hfil}\else{\hfil\twelverm\folio\hfil}\fi}
\let\rawfootnote=\footnote		
\def\footnote#1#2{{\rm\parindent=20pt\singlespace\hang
  \rawfootnote{#1}{\tenrm#2\hfill\vrule height 0pt depth 6pt width 0pt}}}
\def\raggedcenter{\leftskip=4em plus 12em \rightskip=\leftskip
  \parindent=0pt
  \parfillskip=0pt \spaceskip=.3333em \xspaceskip=.5em
  \pretolerance=9999 \tolerance=9999
  \hyphenpenalty=9999 \exhyphenpenalty=9999 }
\hsize=6.5truein
\vsize=8.9truein
\parskip=\medskipamount
\twelvepoint		
\overfullrule=0pt	
\font \titlefont=cmr10 scaled \magstep4
\def \bigtitle{\null\vskip 3pt plus 0.3fill \beginparmode  
           \triplespace \raggedcenter \titlefont}
\def\author			
  {\vskip 3pt plus 0.3fill \beginparmode \raggedcenter \sc}
\def\affil			
  {\vskip 3pt plus 0.1fill \beginlinemode
   \oneandahalfspace \raggedcenter \sl}
\def\abstract			
  {\vskip 3pt plus 0.3fill \beginparmode
   \oneandahalfspace \narrower ABSTRACT:~~}
\def\body{\beginparmode}
\def\subhead#1{\vskip 0.25truein{\raggedcenter #1 \par}
   \nobreak\vskip 0.25truein\nobreak}
\def\references
  {\subhead{REFERENCES}
   \frenchspacing \parindent=0pt \leftskip=0.8truecm \rightskip=0truecm
   \parskip=4pt plus 2pt \everypar{\hangindent=\parindent}}
\def\refstylepr{		
  \gdef\r##1{~[##1]}	         			
  \gdef\refis##1{\indent\hbox to 0pt{\hss[##1]~}}     	
  \gdef\journal##1, ##2, ##3,                           
    ##4{##1 {\bf ##2}, ##3 ##4}}
\def\prd{\journal Phys. Rev. D}
\def\prl{\journal Phys. Rev. Lett.}
\def\cmp{\journal Comm. Math. Phys.}
\def\np{\journal Nucl. Phys.}

\def\endreferences{\body}
\def\endit{\endmode\vfill\supereject\end}
\def\frac#1#2{{\textstyle{#1 \over #2}}}
\def\half{{\textstyle{ 1\over 2}}}
\def\ts{\textstyle}
\def\ds{\displaystyle}

\def\sss{\scriptscriptstyle}
\def\cfpa{Center for Particle Astrophysics\\University of California\\
Berkeley, CA 94720}
\def\lbl{Theoretical Physics Group\\Lawrence Berkeley Laboratory\\
1 Cyclotron Road\\Berkeley, CA 94720}
\def\Tr{\mathop{\rm Tr}\nolimits}
\def\Re{\mathop{\rm Re}\nolimits}
\def\Im{\mathop{\rm Im}\nolimits}

\refstylepr

\catcode`@=11
\newcount\r@fcount \r@fcount=0
\newcount\r@fcurr
\immediate\newwrite\reffile
\newif\ifr@ffile\r@ffilefalse
\def\w@rnwrite#1{\ifr@ffile\immediate\write\reffile{#1}\fi\message{#1}}
\def\writer@f#1>>{}
\def\referencefile{
  \r@ffiletrue\immediate\openout\reffile=\jobname.ref%
  \def\writer@f##1>>{\ifr@ffile\immediate\write\reffile%
    {\noexpand\refis{##1} = \csname r@fnum##1\endcsname = %
     \expandafter\expandafter\expandafter\strip@t\expandafter%
     \meaning\csname r@ftext\csname r@fnum##1\endcsname\endcsname}\fi}%
  \def\strip@t##1>>{}}

\def\citeall#1{\xdef#1##1{#1{\noexpand\cite{##1}}}}
\def\cite#1{\each@rg\citer@nge{#1}}	
\def\each@rg#1#2{{\let\thecsname=#1\expandafter\first@rg#2,\end,}}
\def\first@rg#1,{\thecsname{#1}\apply@rg}	
\def\apply@rg#1,{\ifx\end#1\let\next=\relax
\else,\thecsname{#1}\let\next=\apply@rg\fi\next}
\def\citer@nge#1{\citedor@nge#1-\end-}	
\def\citer@ngeat#1\end-{#1}
\def\citedor@nge#1-#2-{\ifx\end#2\r@featspace#1 
  \else\citel@@p{#1}{#2}\citer@ngeat\fi}	
\def\citel@@p#1#2{\ifnum#1>#2{\errmessage{Reference range #1-#2\space is bad.}%
    \errhelp{If you cite a series of references by the notation M-N, then M and
    N must be integers, and N must be greater than or equal to M.}}\else%
 {\count0=#1\count1=#2\advance\count1
by1\relax\expandafter\r@fcite\the\count0,%
  \loop\advance\count0 by1\relax
    \ifnum\count0<\count1,\expandafter\r@fcite\the\count0,%
  \repeat}\fi}
\def\r@featspace#1#2 {\r@fcite#1#2,}	
\def\r@fcite#1,{\ifuncit@d{#1}
    \newr@f{#1}%
    \expandafter\gdef\csname r@ftext\number\r@fcount\endcsname%
                     {\message{Reference #1 to be supplied.}%
                      \writer@f#1>>#1 to be supplied.\par}%
 \fi%
 \csname r@fnum#1\endcsname}
\def\ifuncit@d#1{\expandafter\ifx\csname r@fnum#1\endcsname\relax}%
\def\newr@f#1{\global\advance\r@fcount by1%
    \expandafter\xdef\csname r@fnum#1\endcsname{\number\r@fcount}}
\let\r@fis=\refis			
\def\refis#1#2#3\par{\ifuncit@d{#1}
   \newr@f{#1}%
   \w@rnwrite{Reference #1=\number\r@fcount\space is not cited up to now.}\fi%
  \expandafter\gdef\csname r@ftext\csname r@fnum#1\endcsname\endcsname%
  {\writer@f#1>>#2#3\par}}
\def\ignoreuncited{
   \def\refis##1##2##3\par{\ifuncit@d{##1}%
     \else\expandafter\gdef\csname r@ftext\csname
r@fnum##1\endcsname\endcsname%
     {\writer@f##1>>##2##3\par}\fi}}
\def\r@ferr{\endreferences\errmessage{I was expecting to see
\noexpand\endreferences before now;  I have inserted it here.}}
\let\r@ferences=\references
\def\references{\r@ferences\def\endmode{\r@ferr\par\endgroup}}
\let\endr@ferences=\endreferences
\def\endreferences{\r@fcurr=0
  {\loop\ifnum\r@fcurr<\r@fcount
    \advance\r@fcurr by 1\relax\expandafter\r@fis\expandafter{\number\r@fcurr}%
    \csname r@ftext\number\r@fcurr\endcsname%
  \repeat}\gdef\r@ferr{}\endr@ferences}

\let\r@fend=\endpaper\gdef\endpaper{\ifr@ffile
\immediate\write16{Cross References written on []\jobname.REF.}\fi\r@fend}
\catcode`@=12
\citeall\r		%

\input epsf
\def\mpl{M_{\rm\sss Pl}}
\def\sbh{S_{\rm\sss BH}}
\def\shr{S_{\rm\sss HR}}
\def\rout{\rho_{\rm out}}
\def\rin{\rho_{\rm in}}
\def\op{\omega_+}
\def\om{\omega_-}
\def\p{\varphi}

\oneandahalfspace
\centerline{March 8, 1993 \hfill CfPA--93--02 \hfill LBL--33754}
\rightline{hep-th/9303048}

\bigtitle Entropy$\;$ and$\;$ Area\footnote{$^{\ds *}$}{This work was
supported in part by the Director, Office of Energy Research, Office of High
Energy and Nuclear Physics, Division of High Energy Physics of the
U.S.~Department of Energy under Contract DE-AC03-76SF00098, and in part by
the National Science Foundation under Grant Nos.~AST91-20005 and PHY91-16964.}

\author Mark Srednicki\footnote{$^\dagger$}{E-mail: mark@tpau.physics.ucsb.edu.
On leave from Department of Physics, University of California, Santa Barbara,
CA 93106.}

\affil\cfpa
\affil{\rm and}
\affil\lbl

\abstract
The ground state density matrix for a massless free field is traced over the
degrees of freedom residing inside an imaginary sphere; the resulting entropy
is shown to be proportional to the area (and not the volume) of the sphere.
Possible connections with the physics of black holes are discussed.

\vfill\eject\beginparmode

\pageno=0
\oneandahalfspace
\phantom{X}
\vskip 1.5in
{\tenrm
This document was prepared as an account of work sponsored by the United
States Government.  Neither the United States Government nor any agency
thereof, nor The Regents of the University of California, nor any of their
employees, makes any warranty, express or implied, or assumes any legal
liability or responsibility for the accuracy, completeness, or usefulness
of any information, apparatus, product, or process disclosed, or represents
that its use would not infringe privately owned rights.  Reference herein
to any specific commercial products process, or service by its trade name,
trademark, manufacturer, or otherwise, does not necessarily constitute or
imply its endorsement, recommendation, or favoring by the United States
Government or any agency thereof, or The Regents of the University of
California.  The views and opinions of authors expressed herein do not
necessarily state or reflect those of the United States Government or any
agency thereof of The Regents of the University of California and shall
not be used for advertising or product endorsement purposes.}
\vskip 2in
\centerline{\it Lawrence Berkeley Laboratory is an equal opportunity employer.}
\vfill\eject
\firstpageno=1
\pageno=1
A free, massless, scalar, quantum field (which could just as well represent,
say, the acoustic modes of a crystal, or any other three-dimensional system
with dispersion relation $\omega=c|\vec k|\,$) is in its nondegenerate
ground state, $|0\rangle$.  We form the ground state density matrix,
$\rho_0=|0\rangle\langle 0|$, and trace over the field degrees of freedom
located inside an imaginary sphere of radius $R$.  The resulting density
matrix, $\rout$, depends only on the degrees of freedom outside the sphere.
We now compute the associated entropy, $S=-\Tr\rout\log\rout$.
How does $S$ depend on $R\,$?

Entropy is usually an extensive quantity, so we might expect that
$S\sim R^3$.  However, this is not likely to be correct, as can be seen
from the following argument.  Consider tracing over the outside degrees of
freedom instead, to produce a density matrix $\rin$ which depends only on
the inside degrees of freedom.  If we now compute $S'=-\Tr\rin\log\rin$,
we would expect that $S'$ scales like the volume {\it outside} the sphere.
However, it is straightforward to show that $\rin$ and $\rout$ have the
same eigenvalues (with extra zeroes for the larger, if they have different
rank), so that in fact $S'=S$\r{eval}.  This indicates that $S$ should
depend only on properties which are shared by the two regions (inside and
outside the sphere).  The one feature they have in common is their shared
boundary, so it is reasonable to expect that $S$ depends only on the area of
this boundary, $A=4\pi R^2$.  $S$ is dimensionless, so to get a nontrivial
dependence of $S$ on $A$ requires another dimensionful parameter.
We have two at hand: the ultraviolet cutoff $M$ and the infrared
cutoff $\mu$, both of which are necessary to give a precise definition
of the theory.  (For a crystal, $M$ would be the inverse atomic
spacing, and $\mu$ the inverse linear size, in units with $\hbar=c=1$.)
Physics in the interior region should be independent of $\mu$, which
indicates that perhaps $S$ will be as well.  We therefore expect that
$S$ is some function of $M^2\!A$.

In fact, as will be shown below, $S=\kappa M^2\!A$, where $\kappa$
is a numerical constant which depends only on the precise definition of
$M$ that we adopt.

This result bears a striking similarity to the formula for the intrinsic
entropy of a black hole, $\sbh=\frac14\mpl^2 A$, where $\mpl$ is the Planck
mass and $A$ is the surface area of the horizon of the black hole\r{sbh}.
The links in the chain of reasoning establishing this formula are remarkably
diverse, involving, in turn, classical geometry, thermodynamic analogies, and
quantum field theory in curved space.  The result is thus rather mysterious.
In particular, we would like to know whether or not $\sbh$ has anything to
do with the number of quantum states accessible to the black hole.

As a black hole evaporates and shrinks, it produces Hawking radiation whose
entropy, $\shr$, can be computed by standard methods of statistical mechanics.
One finds, after the black hole has shrunk to negligible size, that $\shr$ is
a number of order one (depending on the masses and spins of the elementary
particles) times the original black hole entropy\r{shr}.
This calculation of $\shr$ is done by counting quantum states,
and the fact that $\sbh\simeq\shr$ lends support to the idea that $\sbh$
should also be related to a counting of quantum states.  It is then
tempting think of the horizon as a kind of membrane\r{tho}, with
approximately one degree of freedom per Planck area.  However, in
classical general relativity, the horizon does not appear to be
a special place to a nearby observer, so it is hard to see why it should
behave as an object with local dynamics.  The new result quoted above
indicates that $S\sim A$ is a much more general formula than has
heretofore been realized.  It shows that the amount of missing information
represented by $\sbh$ is about right, in the sense that we would get
the same answer in the vacuum of flat space if we did not permit ourselves
access to the interior of a sphere with surface area $A$, and set the
ultraviolet cutoff to be of order $\mpl$ (perfectly reasonable for comparison
with a quantum theory that includes gravity).  Furthermore, getting
$S\sim A$ clearly does not require the boundary of the inaccessible
region to be dynamical, since in our case it is entirely imaginary.

To establish that $S=\kappa M^2\!A$ for the problem at hand, let us
begin with the simplest possible version of it: two coupled harmonic
oscillators, with hamiltonian
$$H = \half\bigl[p_1^2 + p_2^2 + k_0(x_1^2 + x_2^2)
                               + k_1(x_1-x_2)^2\bigr]\;.  \eqno(1)$$
The normalized ground state wave function is
$$\psi_0(x_1,x_2)=\pi^{-1/2}(\op\om)^{1/4}
                  \exp\bigl[-(\op x_+^2 + \om x_-^2)/2\bigr]\;,  \eqno(2)$$
where $x_\pm=(x_1\pm x_2)/\sqrt2$, $\op=k_0^{1/2}$,
and $\om=(k_0+2k_1)^{1/2}$.
We now form the ground state density matrix, and trace over the first
(``inside'') oscillator, resulting in a density matrix for the second
(``outside'') oscillator alone:
$$\eqalign{
\rout(x_2,x'_2) &= \int_{-\infty}^{+\infty} dx_1\,\psi_0(x_1,x_2)
                                                \psi^*_0(x_1,x'_2) \cr
\noalign{\medskip}
                &= \pi^{-1/2}(\gamma-\beta)^{1/2}
       \exp\bigl[-\gamma(x_2^2+x^{\prime 2}_2)/2+\beta x_2x'_2\bigr]\;,\cr}
                                                           \eqno(3)$$
where $\beta=\frac14(\op-\om)^2/(\op+\om)$ and
$\gamma-\beta=2\op\om/(\op+\om)$.
We would like to find the eigenvalues $p_n$ of $\rout(x,x')$:
$$\int_{-\infty}^{+\infty}dx'\,\rout(x,x')f_n(x')=p_n f_n(x)\;,
                                                             \eqno(4)$$
because in terms of them the entropy is simply $S=-\sum_n p_n\log p_n$.
The solution to Eq.~(4) is found most easily by guessing, and is
$$\eqalign{
p_n &= (1-\xi)\xi^n\;,\cr
\noalign{\medskip}
f_n(x) &= H_n(\alpha^{1/2}x)\exp(-\alpha x^2/2)\;, \cr}  \eqno(5)$$
where $H_n$ is a Hermite polynomial,
$\alpha=(\gamma^2-\beta^2)^{1/2}=(\op\om)^{1/2}$,
$\xi=\beta/(\gamma+\alpha)$, and $n$ runs from zero to infinity.
Eq.~(5) imples that $\rout$ is equivalent to a thermal density matrix for a
single harmonic oscillator specified by frequency $\alpha$ and temperature
$T=\alpha/\log(1/\xi)$.  The entropy is
$$S(\xi)=-\log(1-\xi)-{\xi\over 1-\xi}\log\xi \;,\eqno(6)$$
where $\xi$ is ultimately a function only of the ratio $k_1/k_0$.

We can easily expand this analysis to a system of $N$ coupled
harmonic oscillators with hamiltonian
$$H=\half\sum_{i=1}^Np_i^2+\half\sum_{i,j=1}^N x_i K_{ij}x_j\;,  \eqno(7)$$
where $K$ is a real symmetric matrix with positive eigenvalues.  The
normalized ground state wave function is
$$\psi_0(x_1,\ldots,x_N)=\pi^{-N/4}(\det\Omega)^{1/4}
              \exp\bigl[-x\!\cdot\!\Omega\!\cdot\!x/2\bigr]\;,  \eqno(8)$$
where $\Omega$ is the square root of $K$: if $K=U^{\rm T}K_{\rm\sss D}U$,
where $K_{\rm\sss D}$ is diagonal and $U$ is orthogonal, then
$\Omega=U^{\rm T}K^{1/2}_{\rm\sss D}U$.  We now trace over the first $n$
(``inside'') oscillators to get
$$\eqalignno{
\rout(x_{n+1},\ldots,x_N;x'_{n+1},\ldots,x'_N)
&=\int\prod_{i=1}^n dx_i\, \psi_0(x_1,\ldots,x_n,x_{n+1},\ldots,x_N)\cr
&\qquad\qquad\times\psi^*_0(x_1,\ldots,x_n,x'_{n+1},\ldots,x'_N)\;.&(9)\cr}$$
To carry out these integrals explicitly, we write
$$\Omega=\pmatrix{A&B\cr
\noalign{\medskip}
          B^{\rm T}&C\cr}\;, \eqno(10)$$
where $A$ is $n\times n$ and $C$ is $(N-n)\times(N-n)$.  We find
$$\rout(x,x')\sim
  \exp\bigl[-(x\!\cdot\!\gamma\!\cdot\!x+x'\!\!\cdot\!\gamma\!\cdot\! x')/2
             +x\!\cdot\!\beta\!\cdot\! x'\bigr]\;,  \eqno(11)$$
where $x$ now has $N-n$ components, $\beta=\half B^{\rm T}A^{-1}B$, and
$\gamma=C-\beta$.  In general $\beta$ and $\gamma$ will not commute, which
implies that Eq.~(11) is {\it not} equivalent to a thermal density matrix
for a system of oscillators.

We need not keep track of the normalization of $\rout$,
since we know that its eigenvalues must sum to one.  To find them, we note
that the appropriate generalization of Eq.~(4) implies that
$(\det G)\,\rout(Gx,Gx')$ has the same eigenvalues as $\rout(x,x')$, where $G$
is any nonsingular matrix.  Let $\gamma=V^{\rm T}\gamma_{\rm\sss D}V$, where
$\gamma_{\rm\sss D}$ is diagonal and $V$ is orthogonal; then let
$x=V^{\rm T}\gamma^{-1/2}_{\rm\sss D}y$.  (The eigenvalues of $\gamma$ are
guaranteed to be positive, so this transformation is well defined.)
We then have
$$\rout(y,y')\sim\exp\bigl[-(y\!\cdot\! y+y'\!\!\cdot\! y')/2
                     +y\!\cdot\!\beta'\!\!\cdot\! y'\bigr]\;,   \eqno(12)$$

\noindent where
$\beta'=\gamma^{-1/2}_{\rm\sss D}V\beta V^{\rm T}\gamma^{-1/2}_{\rm\sss D}$.
If we now set $y=Wz$, where $W$ is orthogonal and $W^{\rm T}\beta' W$
is diagonal, we get
$$\rout(z,z')\sim\prod_{i=n+1}^N
   \exp\bigl[-(z_i^2+z_i^{\prime 2})/2
             +\beta'_i z^{\phantom\prime}_i z'_i\bigr]\;, \eqno(13)$$
where $\beta'_i$ is an eigenvalue of $\beta'$.
Each term in this product is identical to the $\rout$ of Eq.~(3), with
$\gamma\to 1$ and $\beta\to\beta'_i$.  Therefore, the entropy associated
with the $\rout$ of Eq.~(13) is just $S=\sum_i S(\xi_i)$, where $S(\xi)$
is given by Eq.~(6), and
$\xi_i=\beta'_i\big/\bigl[1+(1-\beta^{\prime 2}_i)^{1/2}\bigr]$.

We now wish to apply this general result to a quantum field with hamiltonian
$$H=\half\int d^3x\,\bigr[\pi^2(\vec x\,)+|\nabla\p(\vec x\,)|^2\bigl]\;.
                                                                 \eqno(14)$$
To regulate this theory, we first introduce the partial wave components
$$\eqalign{
\p_{lm}(x) &= x\int d\Omega\,Z_{lm}(\theta,\phi)\p(\vec x\,)\;,\cr
\pi_{lm}(x) &= x\int d\Omega\,Z_{lm}(\theta,\phi)\pi(\vec x\,)\;,\cr}
                                                               \eqno(15)$$
where $x=|\vec x\,|$ and the $Z_{lm}$ are real spherical harmonics:
$Z_{l0}=Y_{l0}$, $Z_{lm}=\sqrt2\Re Y_{lm}$ for $m>0$, and
$Z_{lm}=\sqrt2\Im Y_{lm}$ for $m<0$; the $Z_{lm}$ are orthonormal
and complete.  The operators defined in Eq.~(15) are hermitian, and
obey the canonical commutation relations
$$\bigl[\p_{lm}(x),\pi_{l'm'}(x')\bigr]
                   =i\delta_{ll'}\delta_{mm'}\delta(x-x')\;.   \eqno(16)$$
In terms of them, we can write $H=\sum_{lm}H_{lm}$, where
$$H_{lm}=\half\int_0^\infty dx\,\left\{\pi^2_{lm}(x)
+ x^2\left[{\partial\over\partial x}\left({\p_{lm}(x)\over x}\right)\right]^2
+ {l(l+1)\over x^2}\,\p_{lm}^2(x) \right\}\;.                     \eqno(17)$$
So far we have made no approximations or regularizations.

Now, as an ultraviolet regulator, we replace the continuous
radial coordinate $x$ by a lattice of discrete points with
spacing $a$; the ultraviolet cutoff $M$ is thus $a^{-1}$.
As an infrared regulator, we put the system in a spherical box of radius
$L=(N+1)a$, where $N$ is a large integer, and demand that $\p_{lm}(x)$ vanish
for $x\ge L$; the infrared cutoff $\mu$ is thus $L^{-1}$.  All together,
this yields
$$H_{lm}={1\over2a}\sum_{j=1}^N\left[\pi^2_{lm,j}
+ (j+\half)^2\left({\p_{lm,j}\over j}-{\p_{lm,j+1}\over j+1}\right)^2
+ {l(l+1)\over j^2}\,\p_{lm,j}^2 \right]\;,                      \eqno(18)$$
where $\p_{lm,N+1}=0$; $\p_{lm,j}$ and $\pi_{lm,j}$ are dimensionless,
hermitian, and  obey the canonical commutation relations
$$\bigl[\p_{lm,j},\pi_{l'm',j'}\bigr]=i\delta_{ll'}\delta_{mm'}\delta_{jj'}\;.
                                                               \eqno(19)$$
Thus, $H_{lm}$ has the general form of Eq.~(7), and for a fixed value of $N$
we can compute (numerically) the entropy $S_{lm}(n,N)$ produced by tracing the
ground state of $H_{lm}$ over the first $n$ sites.  The ground state of $H$
is a direct product of the ground states of each $H_{lm}$, and so the total
entropy is found by summing over $l$ and $m$:  $S(n,N)=\sum_{lm}S_{lm}(n,N)$.
As can be seen from Eq.~(18), $H_{lm}$ is actually independent of $m$,
and therefore so is $S_{lm}(n,N)=S_l(n,N)$.  Summing over $m$ just yields
a factor of $2l+1$, and so we have $S(n,N)=\sum_l(2l+1)S_l(n,N)$.  From
Eq.~(18) we also see that the $l$-dependent term dominates if $l\gg N$,
and in this case we can compute $S_l(n,N)$ perturbatively.  The result is that,
for $l\gg N$, $S_l(n,N)$ is independent of $N$, and is given by
$$S_l(n,N)=\xi_l(n)\bigl[-\log\xi_l(n)+1\bigr]\;,               \eqno(20)$$
where
$$\xi_l(n)={n(n+1)(2n+1)^2\over 64\,l^2(l+1)^2}+O(l^{-6})\;.    \eqno(21)$$
Eqs.~(20) and~(21) demonstrate that the sum over $l$ will converge,
and also provide a useful check on the numerical results.

Let us define $R=(n+\half)a$, a radius midway between the outermost point
which was traced over, and the innermost point which was not.  The computed
values of $S(n,N)$ are shown for $N=60$ and $1\le n\le 30$ as a function
of $R^2$ in Fig.~1.  As can be seen, the points are beautifully fit by
a straight line:
$$S=0.30\,M^2 R^2\;,                                      \eqno(22)$$
where $M=a^{-1}$.  Furthermore, $S(n,N)$ turns out to be independent
of $N$ (and hence the value of the infrared cutoff).  Specifically, for
fixed $n$, with $n\le\half N$, the values of $S(n,N)$ turn out to be
identical (in the worst case, to within 0.5\%) for $N=20$, 40, and~60.
The restriction to $n\le\half N$ is necessary, since the linear behavior
in Fig.~1 cannot continue all the way to $n=N$:  at this point we will
have traced over {\it all} the degrees of freedom, and must find $S=0$.
$S$ must therefore start falling as $R$ begins to approach the wall of
the box at radius $L=(N+1)a$.

Of course, similar calculations can be done for one- and two-dimensional
systems as well.  For $d=2$, our introductory arguments would lead us to
expect that $S=\kappa M\!R$, since the relevant ``area'' is the circumference
of the dividing circle of radius $R$.  This is confirmed by the numerical
results, which will be presented in detail elsewhere\r{inp}.  For $d=1$,
our arguments must break down:  they would lead to the conclusion that
$S$ is independent of $R$, and this is clearly impossible.  In fact, the
numerical results indicate that $S=\kappa_1\log(M\!R)+\kappa_2\log(\mu R)$ in
one dimension; for the first time, we see a dependence on the infrared cutoff
$\mu$\r{sd1}.  For $d\ge4$, regularization by a radial lattice turns out to be
insufficient; the sum over partial waves does not converge.  Regularization by
a full, $d$-dimensional lattice would certainly produce a finite $S$, but this
procedure would greatly increase the computational complexity.

To summarize, a straightforward counting of quantum states in a simple,
well-defined context has produced an entropy proportional to the surface
area of the inaccessible region, inaccessible in the sense that we ignore
the information contained there.  Eq.~(22) is strikingly similar to the
formula for the entropy of a black hole, $\sbh=\frac14\mpl^2 A$,
and so may provide some clues as to its deeper meaning.

I am grateful to Orlando Alvarez, Steve Giddings, David Gross, and Andy
Strominger for helpful discussions.

Note added:  After this paper was completed, I learned of related work
by Bombelli {\it et~al}\r{bomb}.  Also motivated by the black hole analogy,
these authors find an equivalent result for the entropy of a coupled system
of oscillators.  They also argue that, for a quantum field, the entropy
should be proportional to the area of the boundary; the argument they give
is different from those presented here, and is valid only if the field has
a mass $m$ which is large enough to make the Compton wavelength $1/m$
much less than $R$.
I thank Erik Matinez for bringing this paper to my attention.

\vfill\eject

\references
\oneandahalfspace

\refis{eval}Let $|0\rangle
=\sum_{ia}\psi_{ia}|i\rangle_{\rm in}|a\rangle_{\rm\vphantom{in}out}$,
so that $(\rin)_{ij}=(\psi\psi^\dagger)_{ij}$ and
$(\rout)_{ab}=(\psi^{\rm T}\psi^{\ts *})_{ab}$.  Now it is clear that
$\Tr\rin^k=\Tr\rout^k$ for any positive integer~$k$.  This can only be true
if $\rin$ and $\rout$ have the same eigenvalues, up to extra zeroes.

\refis{sbh}J. D. Bekenstein, \prd, 7, 2333, (1973); \prd, 9, 3292, (1974);
S. W. Hawking, \cmp, 43, 199, (1975);
G. W. Gibbons and S. W. Hawking, \prd, 15, 2752, (1977);
S. W. Hawking, in {\sl General Relativity: An Einstein Centenary Survey},
ed. S. W. Hawking and W. Israel (Cambridge, 1979);
R. Kallosh, T. Ort{\'\i}n, and A. Peet, Stanford Univ. Report SU--ITP--92--29
(hep-th/9211015).

\refis{shr}W. H. Zurek, \prl, 49, 1683, (1982);
D. N. Page, \prl, 50, 1013, (1983).

\refis{tho}G. 't Hooft, \np, B355, 138, (1990), and references therein.

\refis{inp}M. Srednicki, in preparation.

\refis{sd1}This formula bears no apparent relation to that for the entropy
of a one-dimensional black hole; see
G. W. Gibbons and M. J. Perry, \journal Int. J. Mod. Phys., D1, 335, (1992);
C. Nappi and A. Pasquinucci, \journal Mod. Phys. Lett., A7, 3337, (1992).

\refis{bomb}L. Bombelli, R. K. Koul, J. Lee, and R. Sorkin, \prd, 34, 373,
(1986).

\endreferences
\vfill\eject

\centerline{\epsfbox{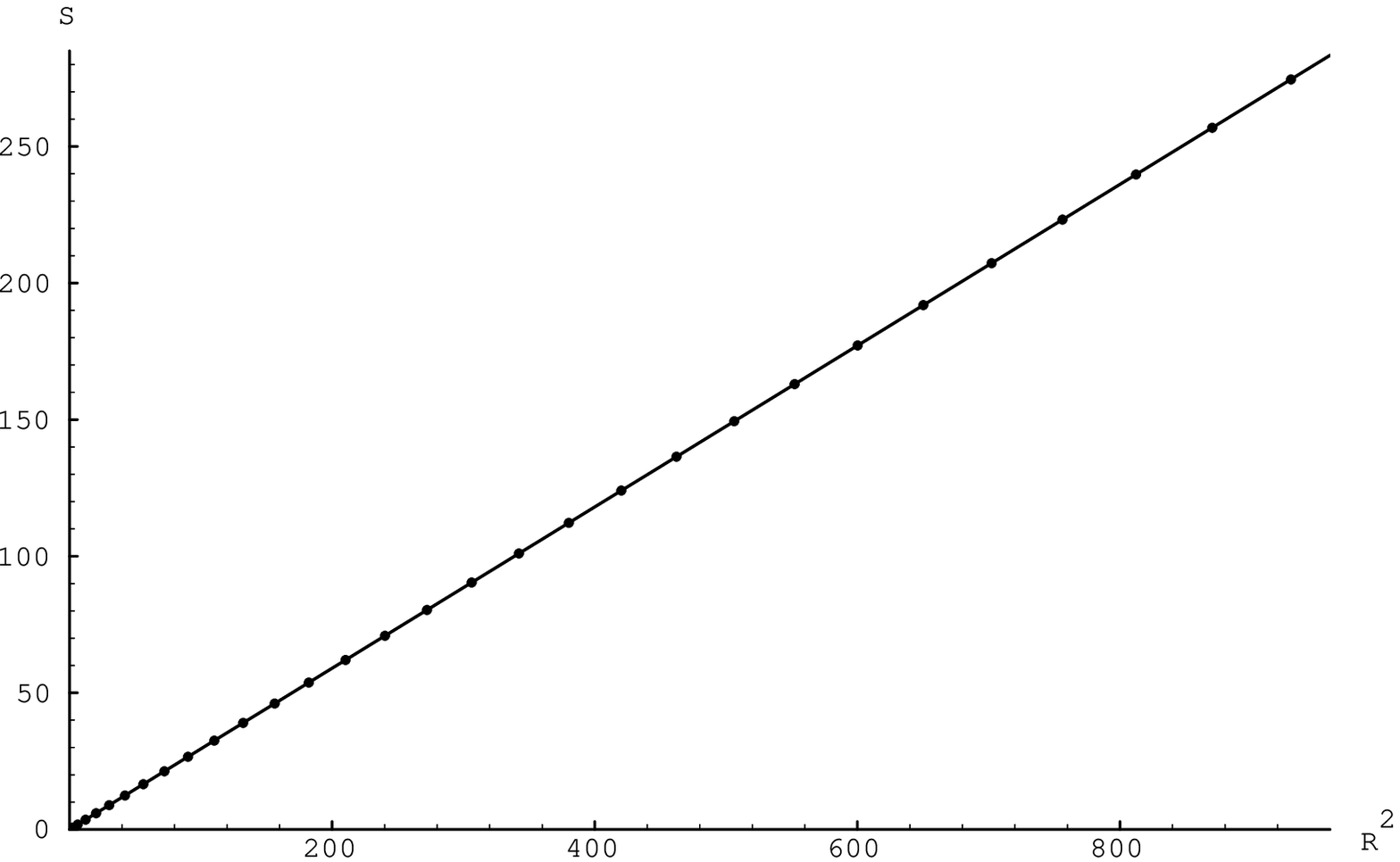}}

\vskip0.8in
\oneandahalfspace

\item{Fig.~1.}The entropy $S$ resulting from tracing the ground state of a
massless scalar field over the degrees of freedom inside a sphere of
radius $R$.  The points shown correspond to regularization by a radial
lattice with $N=60$ sites; the line is the best linear fit.  $R$ is
measured in lattice units, and is defined to be $n+\half$, where $n$ is
the number of traced sites.

\endit\end